\documentclass[12pt]{article}
\usepackage[utf8]{inputenc}
\usepackage{amsmath,amssymb,latexsym}
\usepackage[font=small,labelfont=bf,   justification=justified,   format=plain]{caption}
\usepackage{subcaption}
\usepackage{adjustbox}
\usepackage{enumerate}
\usepackage{graphics}
\usepackage{graphicx}
\usepackage{epsfig}
\usepackage{tikz}
\usepackage[square,sort&compress,comma,numbers]{natbib}
\usepackage{float}
\usepackage{fancyhdr}
\usepackage[a4paper, total={6in, 8in}]{geometry}
\usepackage{booktabs}
\usepackage{makeidx}
\usepackage{amssymb}
\usepackage{eurosym}
\usepackage{placeins}
\usepackage{url}
\usepackage{amsthm}
\usepackage{times}
\usepackage[T1]{fontenc}
\usepackage[utf8]{inputenc}
\usepackage{grffile}
\usepackage{multirow}
\usepackage{multicol}
\graphicspath{{images/}}
\floatstyle{plaintop}
\restylefloat{table}

\usepackage{xcolor}

\newlength\imagewidth
\newlength\imagescale

\emergencystretch=\maxdimen
\begin{document}
\newgeometry{top=3cm,bottom=2cm}  
\title{\bf Geometric Optimization and IPA-Induced Dispersion Tuning in Solid-Core Photonic Crystal Fibers}
\author{Zekeriya Mehmet Yuksel$^{1,6}$\footnote{zyuksel08@posta.pau.edu.tr}, Hasan Oguz$^{2,6}$, Ozgur Onder Karakilinc$^{3}$,\\ Halil Berberoglu$^{4}$,  Mirbek Turduev$^5$, Muzaffer Adak$^{1}$, Sevgi Ozdemir Kart$^{1,6}$\footnote{ozsev@pau.edu.tr}.\\
  {\small $^1$ Department of Physics, Science Faculty,  }\\
  {\small  Pamukkale University, 20160 Kınıklı Campus, Denizli, Türkiye}\\
  {\small $^2$ Department of Computer Technologies, Vocational School,  }\\
  {\small  Istanbul Okan University, 34959 Tuzla, Istanbul, Türkiye }\\
  {\small $^3$Electrical and Electronics Engineering, Faculty of Engineering,}\\
  {\small  Pamukkale University, 20160 Kınıklı Campus, Denizli, Türkiye} \\
  {\small $^4$ Department of Physics, Polatli Faculty of Science and Letters, } \\
  {\small Ankara Haci Bayram Veli University, 06900 Polatli, Ankara, Türkiye }\\
  {\small $^5$ Electrical and Electronics Engineering, Faculty of Engineering, }\\
  {\small Kyrgyz-Turkish Manas University, Bishkek, Chuy, 720038 Kyrgyz Republic }\\
  {\small $^6$ Material Physics Simulation Laboratory, Science Faculty, }\\
  {\small  Pamukkale University, 20160 Pamukkale, Denizli, Turkey}\\
}
  \vskip 0.5cm
\date{}
\maketitle
 \thispagestyle{empty}

\begin{abstract}
\noindent
\small
This study presents a comprehensive numerical investigation of solid-core photonic crystal fibers (PCFs) with circular and hexagonal cladding geometries, aiming to optimize key optical parameters for nonlinear photonics and environmental sensing applications. Full-vectorial simulations using FDTD (Lumerical), PWE (MPB), and FDE (MODE) are employed to analyze the influence of structural parameters—core diameter ($d_c$), pitch ($\Lambda$), and air filling fraction—on zero-dispersion wavelength (ZDW), nonlinear coefficient ($\gamma$), effective mode area ($A_{\text{eff}}$), and confinement loss. The results reveal that decreasing $d_c$ from 2.4\,µm to 1.4\,µm enables ZDW tuning from 791\,nm to 646\,nm, alongside a 72\% increase in $\gamma$, from 72\,W$^{-1}$km$^{-1}$ to 124\,W$^{-1}$km$^{-1}$. The impact of isopropyl alcohol (IPA) infiltration is also examined, demonstrating a significant red-shift in ZDW and reduced index contrast that deteriorates confinement and dispersion slope. These findings establish a robust design framework for PCFs that combines high nonlinear efficiency with resilience against contamination, offering valuable guidance for supercontinuum generation and chemical sensing applications.

\noindent
{\it Keywords}: Photonic crystal fiber, fused Silica fibers, computational photonics

\end{abstract}

\restoregeometry

\section{Introduction}

Photonic crystal fibers (PCFs), a groundbreaking development in optical fibers, are based on the unique properties of photonic crystals. In contrast to traditional optical fibers, PCFs can confine light in hollow cores or solid cores, through unique confinement characteristics, offering innovative waveguiding properties with low loss \cite{Knight:96, Russell:2003}. These fibers can be fabricated to achieve large normal group velocity dispersion, enabling loss compensation without silica doping and its associated increased losses \cite{Mogilevtsev:98, Birks:97}. Such properties allow PCFs to facilitate nonlinear effects like visible continuum and supercontinuum generation \cite{Ranka:00, Dudley:06}. PCFs are wavelength-scaled periodic structures extended along the fiber direction. This structure can be created with many topologies and materials and guide light with various light-guiding mechanisms which is index and gap guiding. These mechanisms can cover a wide frequency range extending to terahertz regions \cite{Cerqueira:2010}.

PCFs' versatility is exemplified by their ability to exploit the photonic bandgap effect or support only the fundamental guided mode within the transparency window of silica \cite{Joannopoulos2008, Knight:98}. This adaptability renders PCFs invaluable in diverse applications such as fiber-optic communications, fiber lasers, and nonlinear devices \cite{Russell:2003, Birks:99}.

Structurally, PCFs are characterized by a cross-section of one or more materials, typically arranged periodically to form the cladding surrounding the core(s) \cite{Russell:2003, Knight:98}. The pioneering PCFs, developed by Philip Russell, feature a hexagonal lattice of air holes in a silica matrix, either with a solid or hollow core \cite{Russell:2003}. Recent innovations have led to varied structural designs, such as Bragg fibers and spiral designs, to manipulate optical properties like birefringence and relative refractive index or use of nonlinear materials for exploiting kerr effect \cite{Birks:97, Cregan:99, Agrawal:13,altai2025}. Recent developments in additive manufacturing technologies have opened up exciting possibilities for creating complex geometric-shaped fibers \cite{golebiewski2024,xu2023}.

The categorization of PCFs is based on their light confinement mechanism, namely index guiding and photonic bandgap \cite{Russell:2003, Joannopoulos2008}. Index-guiding PCFs have a core with a higher average refractive index than the cladding, typically achieved with air holes in the cladding material. This leads to strong confinement suitable for nonlinear optical devices. This guiding is based on total internal reflection (TIR) like traditional fibers with lower loss \cite{Kuhlmey:02}. Photonic bandgap PCFs, conversely, utilize microstructured cladding to create a photonic bandgap, enabling light confinement in lower-index or hollow cores \cite{broeng:99, bjarklev2003photonic, Smith2003}.

Bandgap guiding in PCFs is based on the photonic bandgap effect, where the photonic crystal structure creates a range of wavelengths that cannot propagate through the crystal. This is achieved by arranging air holes in a regular pattern around the fiber core, allowing for a lower refractive index core. Hollow-core photonic crystal fibers (HC-PCFs) differ from traditional fibers in guiding mechanism. However, conventional fibers generally exhibit lower transmission loss in standard telecom windows compared to HC-PCFs \cite{Knight:98.2, Barkou:99, Broeng:99.2, Laude:05}.

Index-guiding (IG) PCFs guide light along the fiber core because it cannot cross into the lower-index cladding due to total internal reflection (TIR), similar to standard optical fibers. IG PCFs are widely used in applications requiring control over dispersion and nonlinearity, such as supercontinuum generation or gas sensors. Hybrid PCFs incorporate both guiding mechanisms simultaneously \cite{Birks:97, Hansen:01, Mou:20, Amin:21, Cerqueira:06}.

The key difference lies in the mechanism of light confinement. Bandgap guiding uses a photonic bandgap for confinement, allowing for unique core compositions like lower-index or hollow cores. In contrast, index guiding relies on a higher-index core to confine light through TIR. Hybrid PCFs integrate both modes, offering versatile solutions for various applications. PCFs are an active research subject with promising applications in biosensing, gas sensing, highly sensitive hydrocarbon sensors, quantum dots, signal multiplexing, and supercontinuum generation \cite{Cerqueira:2010, Mittal:2021, Chaudhary:2023,Chen2024, mohsin2025}.

Isopropyl alcohol (IPA) is commonly used in the cleaning and preparation processes of optical fibers, including PCFs. However, IPA contamination can significantly affect the performance and reliability of these fibers. IPA can easily enter the air holes of PCFs due to capillary action, leading to changes in the refractive index and, consequently, the optical properties of the fibers. This contamination can degrade the confinement of light within the core, increase losses, and affect the dispersion characteristics, which are critical for applications such as sensing and communications.

IPA contamination can lead to long-term effects on the performance of PCFs, including increased attenuation and altered nonlinear optical properties. Understanding and mitigating these effects are crucial for the development of high-performance PCFs, especially in environments where exposure to cleaning solvents is commonly used \cite{portosi2019, Rifat:2016}.

This study aims to enhance and optimize the design parameters of both solid-core hexagonal and circular IG-PCFs for nonlinear optics specifically supercontinuum generation. By thoroughly investigating various design aspects, such as the geometrical structure, material composition, and structural parameters, we seek to maximize the performance of these PCFs and examine the effects of IPA contamination on these structures. Our research focuses on determining the critical optical properties, including the effective index ($n_{eff}$), mode profiles, zero-dispersion wavelength (ZDW), nonlinear coefficient, and effective mode area. The findings will contribute to optimizing the use of PCFs for some other applications like optical communication and sensor technology, advancing the development of high-performance PCFs.

\vspace{1em}
\noindent\textbf{Key Contributions.} This work demonstrates structural tuning of the zero-dispersion wavelength (ZDW) over a 145\,nm range (646--791\,nm) through controlled variation of the core diameter ($d_c$). It quantifies a 72\% enhancement in the nonlinear coefficient $\gamma$ as $d_c$ is reduced from 2.4\,µm to 1.4\,µm, thereby improving the potential for efficient supercontinuum generation. The study compares hexagonal and circular PCFs, highlighting how geometric symmetry impacts mode confinement and dispersion slope. Furthermore, it simulates contamination scenarios involving isopropyl alcohol (IPA) infiltration, which causes measurable red-shifts in ZDW and informs practical sensor designs operating under real-world environmental exposure. Lastly, it validates simulation results via strong agreement with experimental benchmarks, reinforcing the reliability of the combined FDTD, PWE, and FDE numerical workflow.

\section{Methodology}

To evaluate and optimize the optical behavior of solid-core photonic crystal fibers (PCFs), we employed a hybrid computational approach combining three well-established numerical methods: Plane Wave Expansion (PWE), Finite-Difference Time-Domain (FDTD), and Finite-Difference Eigenmode (FDE) analysis. The PWE method was implemented using the MIT Photonic Bands (MPB) package to extract band diagrams and group-velocity dispersion characteristics. For time-domain simulations, both MEEP and Lumerical FDTD solvers were utilized to compute effective indices, attenuation spectra, and Poynting vector fields under Gaussian excitation. The FDE analysis was carried out using Lumerical MODE Solutions to evaluate modal profiles, effective mode area $A_{\text{eff}}$, and the nonlinear coefficient $\gamma$.

The PCF designs considered in this study include both hexagonal and circular air-hole geometries. The geometric parameters—core diameter ($d_c$), pitch ($\Lambda$), and air-filling ratio ($f$)—were varied systematically to probe their influence on dispersion, attenuation, and nonlinear performance. The air-filling fraction was defined as $f_c = 2r/\Lambda$ for circular holes and $f_h = s/\Lambda$ for hexagonal holes, where $r$ and $s$ denote the radius and side length of the air holes, respectively. To ensure reproducibility and parametric flexibility, the geometry generation was fully automated via scripting.

FDTD simulations employed a mesh resolution of 5\,nm and applied perfectly matched layers (PML) with a thickness of 1\,$\mu$m in all directions. Modal excitations were introduced using a fundamental Gaussian mode, and convergence was defined by the condition $\Delta n_{\text{eff}} < 10^{-6}$ between successive iterations. The FDE analysis used a 10\,nm grid step with a conformal meshing algorithm enabled. The simulations were run across the 700–1600\,nm spectral range to capture broadband dispersion shifts and mode confinement behavior. The effective mode area was computed from the transverse electric field $\boldsymbol{E}_t$ using the expression
\[
A_{\rm eff} = \frac{\left| \iint |\boldsymbol{E}_t|^2 \, dx\,dy \right|^2}{\iint |\boldsymbol{E}_t|^4 \, dx\,dy},
\]
and the nonlinear coefficient $\gamma$ was obtained using
\[
\gamma = \frac{2\pi n_2}{\lambda A_{\rm eff}},
\]
where $n_2 = 2.6 \times 10^{-20} \, \text{m}^2/\text{W}$ is the Kerr coefficient of silica.

To quantify the optical dispersion properties, we distinguished between material and waveguide contributions. Material dispersion $D_m$ was evaluated using the Sellmeier equation:
\[
n^2(\lambda) = 1 + \frac{A_1 \lambda^2}{\lambda^2 - B_1} + \frac{A_2 \lambda^2}{\lambda^2 - B_2} + \frac{A_3 \lambda^2}{\lambda^2 - B_3},
\]
and the corresponding dispersion parameter is
\[
D_m = -\frac{\lambda}{c} \frac{d^2 n(\lambda)}{d \lambda^2}.
\]
The waveguide dispersion $D_w$ was computed via the real part of the effective index as
\[
D_w = -\frac{\lambda}{c} \frac{d^2 \text{Re}[n_{\text{eff}}]}{d \lambda^2},
\]
allowing us to write the total dispersion $D$ as $D = D_m + D_w$. For silica-based PCFs at telecom wavelengths, $D_m$ is typically small and can be neglected, leading to $D \approx D_w$. The zero-dispersion wavelength $\lambda_0$ was located by identifying the root of $D(\lambda)$, while the dispersion slope at $\lambda_0$ was evaluated as
\[
S(\lambda_0) = \left. \frac{dD}{d\lambda} \right|_{\lambda = \lambda_0}.
\]

Confinement loss $L_c$ was computed from the imaginary part of the effective index:
\[
L_c = \frac{40\pi}{\ln 10 \, \lambda} \text{Im}[n_{\text{eff}}],
\]
while the total attenuation in dB/km was calculated by
\[
\alpha = \frac{10}{L} \log_{10} \left( \frac{P_{\text{in}}}{P_{\text{out}}} \right).
\]
We also determined the power confinement fraction $\eta$ in the core region as
\[
\eta = \frac{\iint_{\text{core}} S_z \, dA}{\iint_{\text{total}} S_z \, dA},
\]
where $S_z$ is the axial component of the time-averaged Poynting vector.

The nonlinear coefficient $\gamma$ was also derived from the nonlinear Schrödinger equation:
\[
\frac{\partial A}{\partial z} + \beta_1 \frac{\partial A}{\partial t} - i \frac{\beta_2}{2} \frac{\partial^2 A}{\partial t^2} = -i \gamma |A|^2 A,
\]
which provides a full vectorial representation but is computationally expensive. In high-contrast PCFs, the area-based definition remains a practical and accurate alternative.

\begin{figure}[htbp!]
    \centering
        \includegraphics[width=\textwidth]{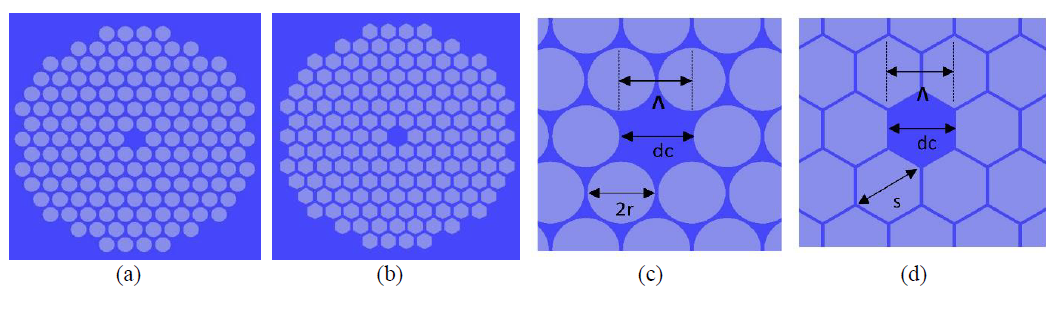} 
        \caption{Schematic cross-sectional view of the proposed PCFs (a) circular air holes composed of PCF-1 and (b) hexagonal air holes composed of PCF-2. Structural parameters of (c) PCF-1 and (d) PCF-2, where $\Lambda$ is the pitch, $d_c$ is the core diameter, $2r$ is the diameter of the circular air holes, and $s$ is the diameter of hexagonal air holes.}
   \label{fig: PCFpara}
\end{figure}

The design of the PCF structure was guided by a benchmark comparison against the Thorlabs NL-2.3-790-02 fiber\cite{thorlabs:pcf}, which has a nominal core diameter of $2.3 \pm 0.1 \,\mu$m. Although the pitch value in the fabricated fiber is 1.6\,$\mu$m, our simulated designs varied $\Lambda$ up to 2.3\,$\mu$m for parameter exploration. Two lattice types were considered: PCF-1 with circular air holes and PCF-2 with hexagonal holes. Design parameters were optimized through multiple FDTD and FDE simulations, targeting enhancements in dispersion, confinement, nonlinearity, and loss performance. A schematic view of both PCF types is presented in Fig.~\ref{fig: PCFpara}, along with definitions of all geometrical parameters.

To validate our model, we compared simulated optical and physical parameters of PCF-2 to experimental data from the Thorlabs fiber. Despite the variation in pitch, our model produced values for zero-dispersion wavelength, nonlinear area, and $\gamma$ that differed by less than 4\% from the experimental benchmarks. Table~\ref{table:comparison_results} summarizes these comparisons and confirms the fidelity of the numerical model. The consistency across parameters such as $A_{\rm eff}$, $S(\lambda_0)$, and $\lambda_0$ demonstrates that the structural modifications introduced during simulation did not compromise the accuracy of the predicted optical response.

\begin{table}[htbp!]
\centering
\caption{Calculated physical and optical properties of the proposed PCF-2, along with the comparison of experimental results}
\label{table:comparison_results}
\begin{tabular}{lcc}
\toprule
\textbf{Properties} & \textbf{Expriment}\textbf{\cite{thorlabs:pcf}} & This Study (\textbf{PCF-2}) \\
\midrule
\textbf{Physical Properties} & & \\
Core Diameter($d_c$) & $2.3 \pm 0.1 \mu m$ & $2.4 \mu m$ \\
Pitch($\Lambda$) & $1.6 \mu m$ & $2.3 \mu m$ \\
Air Filling Fraction & $>94\%$ & $95.83\%$ \\
PCF Region Diameter & $35 \mu m$ & $29.6 \mu m$ \\
\midrule
\textbf{Optical Properties} & & \\
Zero Dispersion Wavelength ($\lambda_0$) & $790 \pm 5$ nm & $787 \pm 10$ nm \\
Dispersion Slope at $\lambda_0$ ($S(\lambda_0)$) & $0.64 \, \text{ps} \cdot \text{nm}^{-1} \cdot \text{km}^{-1}$ & $0.61 \, \text{ps} \cdot \text{nm}^{-1} \cdot \text{km}^{-1}$ \\
Effective Nonlinear Area ($A_{eff}$) & $2.70 \, \mu \text{m}^2$ & $2.75 \, \mu \text{m}^2$ \\
Nonlinear Coefficient at $\lambda_0$ ($\gamma$) & $75 \, \text{W}^{-1} \cdot \text{km}^{-1}$ & $72 \, \text{W}^{-1} \cdot \text{km}^{-1}$ \\
\bottomrule
\end{tabular}
\end{table}

Finally, we also investigated the impact of contamination by isopropyl alcohol (IPA), modeled by assigning the refractive index of the cladding holes as $n_{\text{IPA}} = 1.377$ at 1.55\,$\mu$m. The presence of IPA caused a red-shift in the zero-dispersion wavelength and a measurable increase in loss, highlighting the sensitivity of PCFs to ambient contamination. This aspect is particularly relevant for biosensing and supercontinuum applications, where precise control over modal and spectral properties is essential.

Altogether, the integrated methodology presented here enables a detailed and predictive characterization of PCF behavior, providing a robust foundation for tailoring fiber geometries to specific optical functions in telecommunications, sensing, and nonlinear photonics.

\FloatBarrier

\section{Evaluation of Optical Properties}

We present a comprehensive evaluation of how the core diameter \(d_c\) influences the optical characteristics of photonic crystal fibers (PCFs). Beginning with a baseline structure featuring a core diameter of \(2.4\,\mu\text{m}\)—chosen to closely match the experimentally characterized Thorlabs NL-2.3-790-02 fiber—we systematically reduced \(d_c\) to \(1.4\,\mu\text{m}\) in steps of \(0.2\,\mu\text{m}\). The effects of these dimensional variations were quantified in terms of zero-dispersion wavelength \(\lambda_0\), confinement loss \(L_c\), dispersion slope \(S(\lambda_0)\), attenuation \(\alpha(\lambda_0)\), nonlinear coefficient \(\gamma(\lambda_0)\), and effective mode area \(A_{\text{eff}}\). In addition, we evaluated the impact of isopropyl alcohol (IPA) infiltration, simulating contamination scenarios relevant to biosensing and environmental monitoring.

\begin{figure}[htbp!]
\centering
\includegraphics[width=\textwidth]{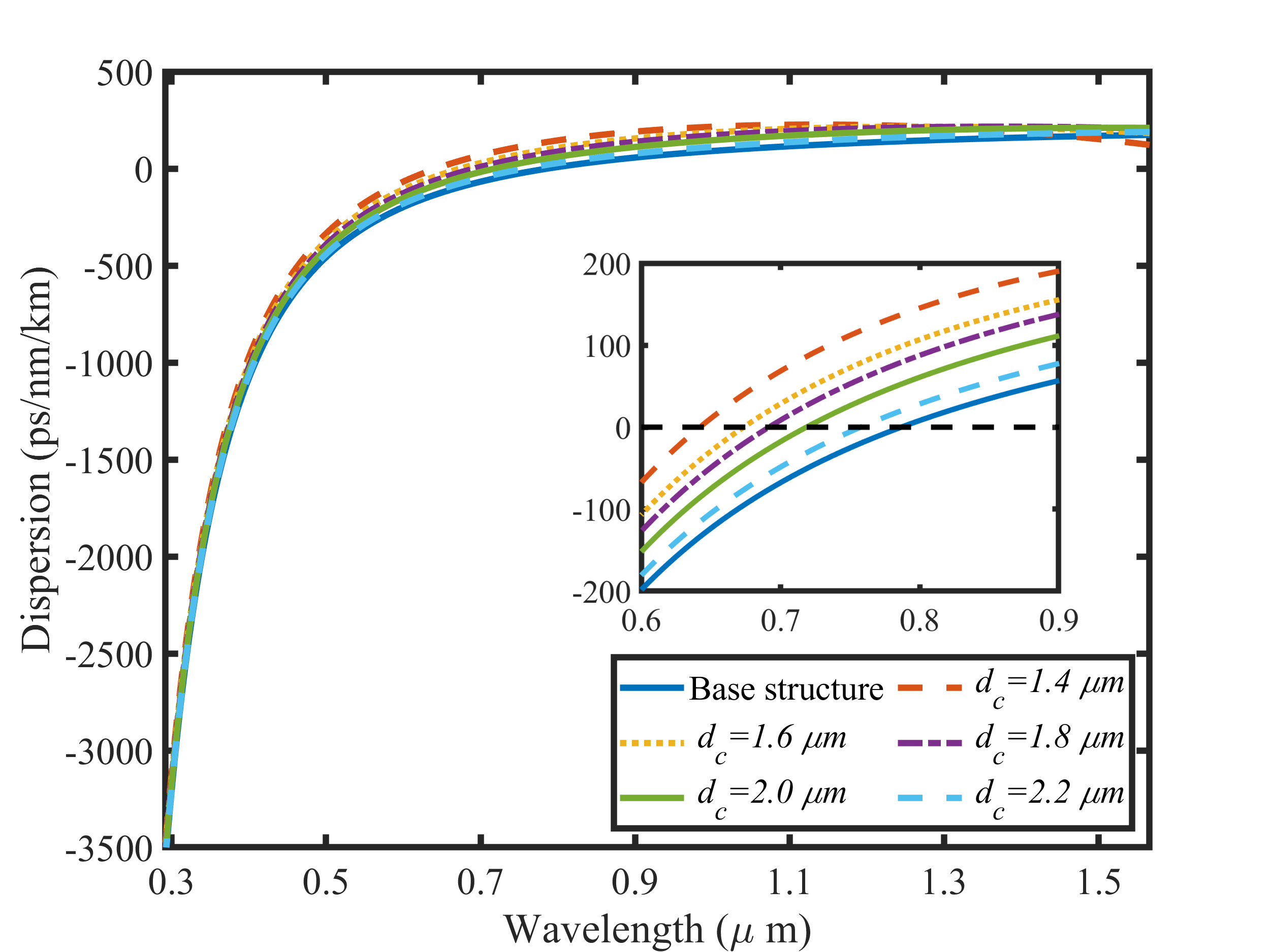}
\caption{Dispersion behavior and ZDW for $d_c = 1.4 \mu m$ to $d_c = 2.2 \mu m$ with $0.2 \mu m$ increase compared to the base structure of PCF-1. Inset showing ZDW for different $d_c$ values.}
\label{fig:disp}
\end{figure}

As illustrated in Fig.~\ref{fig:disp}, reducing \(d_c\) results in a linear blue-shift in \(\lambda_0\), a behavior that allows for precise tailoring of the dispersion characteristics. The inset highlights the tunability of ZDW across the studied range, which varies from \(791\,\text{nm}\) for \(d_c = 2.4\,\mu\text{m}\) to \(646\,\text{nm}\) for \(d_c = 1.4\,\mu\text{m}\) (Table~\ref{table:dispersion_values}). This tunability is instrumental for designing PCFs for specific spectral bands.

\begin{table}[htbp!]
\centering
\caption{Dispersion Values for Different Diameters}
\label{table:dispersion_values}
\begin{tabular}{cc}
\toprule
\textbf{ Diameter $d_c$ ($\mu m$)} & \textbf{Zero Dispersion Wavelength $\lambda_0$ ($nm$)} \\
\midrule
1.4 & 646 \\
1.6 & 676 \\
1.8 & 705 \\
2.0 & 738 \\
2.2 & 758 \\
2.4 & 791 \\
\bottomrule
\end{tabular}
\end{table}

\begin{figure}[htbp!]
\centering
\includegraphics[width=\textwidth]{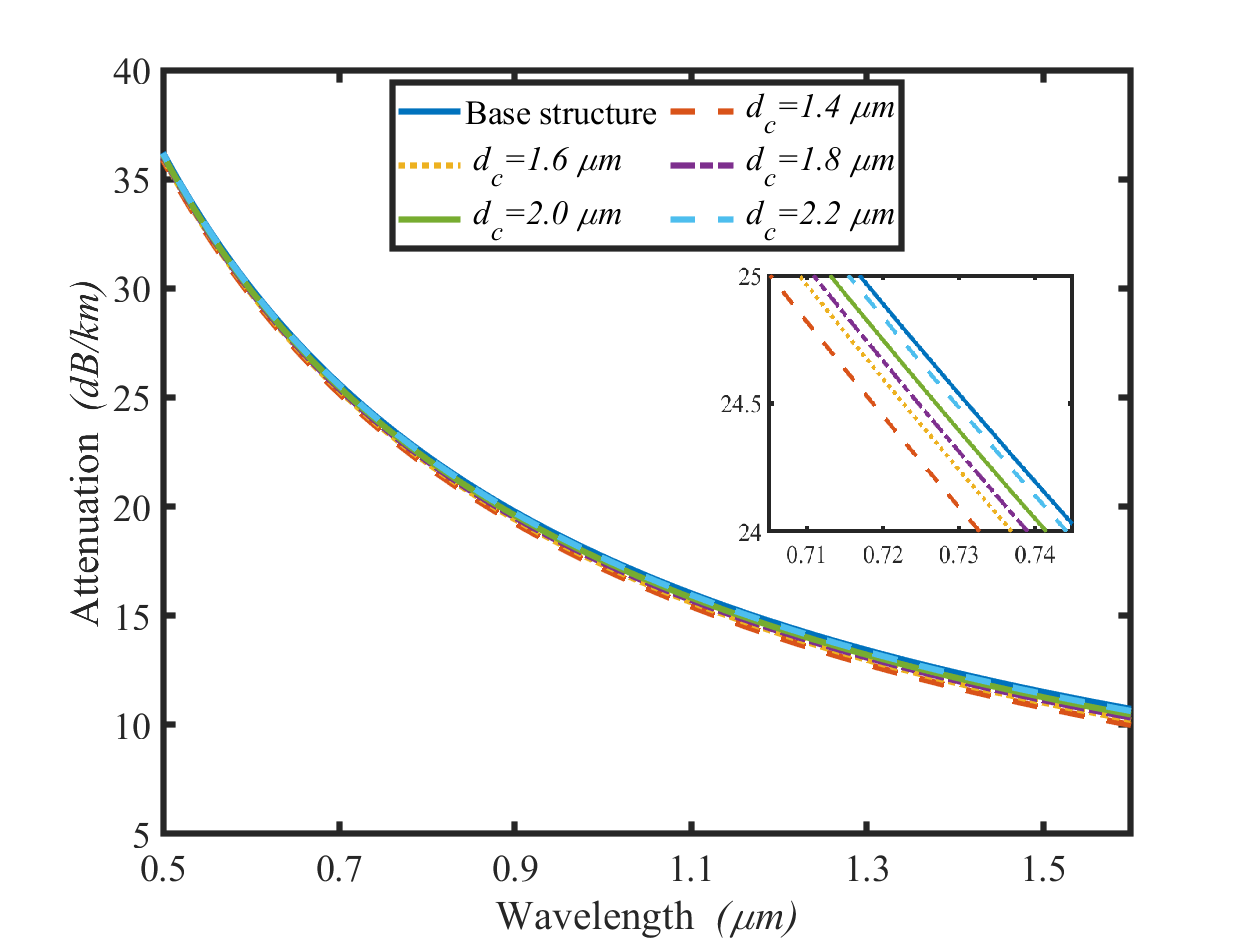}
\caption{Attenuation $L_c$ behavior for $d_c = 1.4 \mu m$ to $d_c = 2.2 \mu m$ with $0.2 \mu m$ increase compared to the base structure of PCF-1. Inset showing $L_c$ of the ZDW region for different $d_c$ values.}
\label{fig: PCFlcall}
\end{figure}

Fig.~\ref{fig: PCFlcall} shows the associated confinement loss profiles. As expected, smaller core diameters increase \(L_c\), reflecting the trade-off between tighter confinement and enhanced dispersion control. For example, \(L_c\) increases from \(0.04\,\text{dB/m}\) at \(d_c = 2.4\,\mu\text{m}\) to \(0.21\,\text{dB/m}\) at \(1.4\,\mu\text{m}\). These values (Table~\ref{table:attenuation}) underscore the importance of balancing confinement with other performance metrics.

\begin{table}[htbp!]
\centering
\caption{Attenuation ($\alpha$) and confinement loss ($L_c$) values for PCF-1 with varying core diameter ($d_c$)}
\label{table:attenuation}
\begin{tabular}{ccc}
\toprule
\textbf{Core Diameter} ($d_c$) & \textbf{Attenuation} $\alpha$ [dB/km] & \textbf{Confinement Loss} $L_c$ [dB/m] \\
\midrule
$2.4\,\mu$m & 22.3 & 0.04 \\
$2.2\,\mu$m & 24.6 & 0.06 \\
$2.0\,\mu$m & 26.9 & 0.09 \\
$1.8\,\mu$m & 28.5 & 0.13 \\
$1.6\,\mu$m & 30.1 & 0.17 \\
$1.4\,\mu$m & 32.7 & 0.21 \\
\bottomrule
\end{tabular}
\end{table}

\begin{figure}[htbp]
\centering
\includegraphics[width=\textwidth]{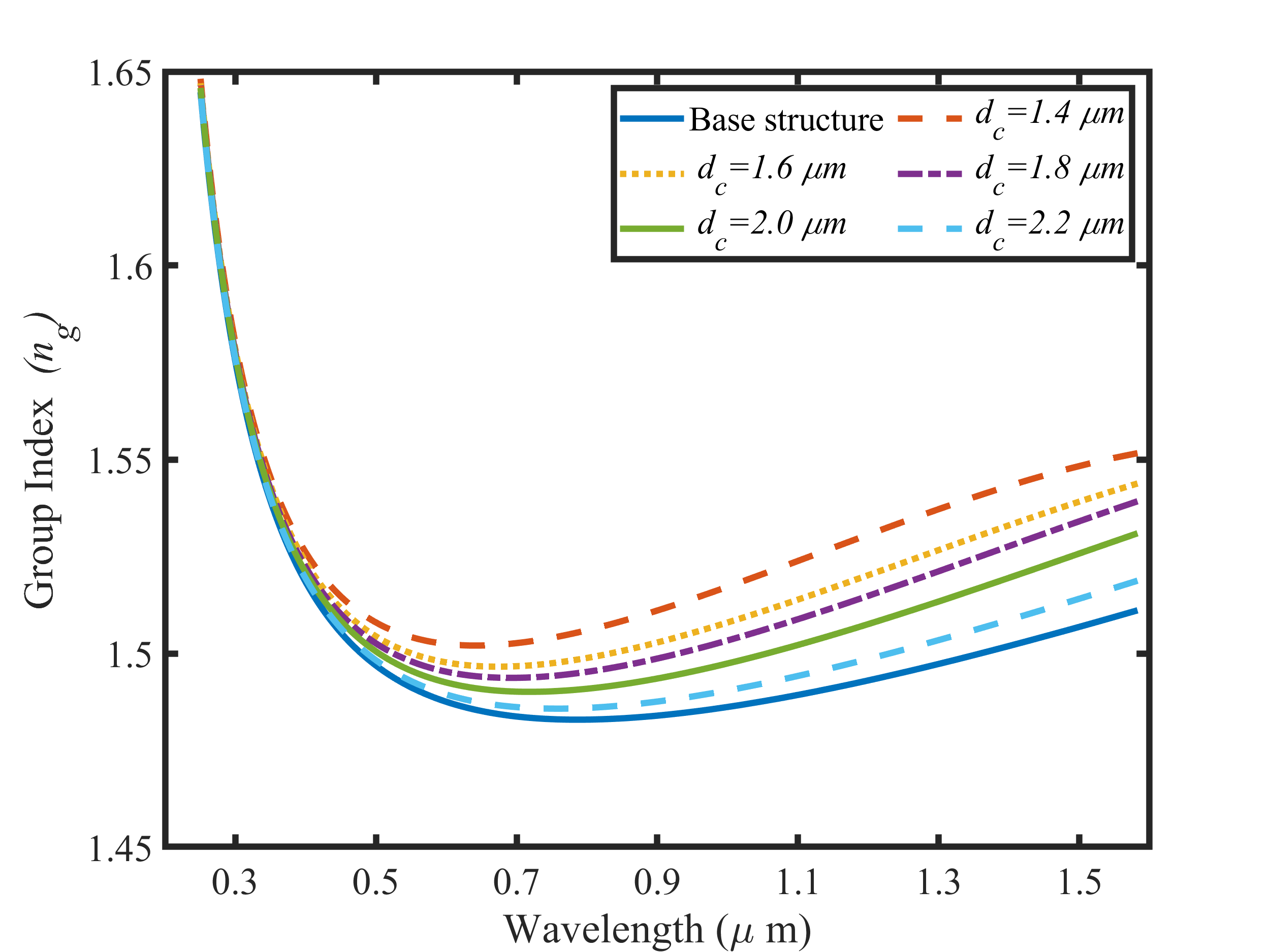}
\caption{Group index $n_g$ behavior for $d_c = 1.4 \mu m$ to $d_c = 2.2 \mu m$ with $0.2 \mu m$ increase compared to the base structure of PCF-1.}
\label{fig: PCFngall}
\end{figure}

Group index \(n_g\) trends, shown in Fig.~\ref{fig: PCFngall}, reveal that higher \(n_g\) is obtained for smaller cores, particularly at longer wavelengths. At \(1.5\,\mu\text{m}\), the group index shifts from \(1.51\) for \(d_c = 2.4\,\mu\text{m}\) to \(1.57\) for \(1.4\,\mu\text{m}\). This has implications for slow-light and dispersion-compensating devices. Moreover, as \(d_c\) decreases and \(n_g\) increases, confinement loss and attenuation also rise, introducing practical trade-offs.

To validate the simulation framework, Table~\ref{table:PCFEXPSIM} compares our simulated results for PCF-1 and PCF-2 against the Thorlabs benchmark. The excellent agreement in \(\lambda_0\), \(S(\lambda_0)\), \(A_{\text{eff}}\), and \(\gamma(\lambda_0)\) confirms the physical fidelity of the model. Notably, PCF-2 closely matches the experimental \(\gamma = 75\,\text{W}^{-1}\text{km}^{-1}\) with a simulated value of \(74.38\,\text{W}^{-1}\text{km}^{-1}\), despite a minor increase in \(A_{\text{eff}}\).

\begin{table}[htbp!]
\centering
\caption{Comparison of Experimental Data \cite{thorlabs:pcf}  and Simulation Results for PCF-1 and PCF-2 Structures}
\label{table:PCFEXPSIM}
\begin{tabular}{llll}
\toprule
 & \cite{thorlabs:pcf} & PCF-1 & PCF-2 \\
\midrule
$d_c$ & $2.30$ $\mu m$ & $2.45$ $\mu m$ & $2.40$ $\mu m$ \\
$\Lambda$ & $1.60$ $\mu m$ & $2.30$ $\mu m$ & $2.00$ $\mu m$ \\
$\lambda_0$ & $790$ $nm$ & $794$ $nm$ & $791$ $nm$ \\
$S(\lambda_0)$ & $0.640$ $ps.nm{-2}.km^{-1}$ & $0.581$ $ps.nm^{-2}.km^{-1}$ & $0.633$ $ps.nm^{-2}.km^{-1}$ \\
$\alpha(\lambda_0)$ & $25$ $dB/km$ & $24.86$ $dB/km$ & $29.67$ $dB/km$ \\
$\gamma_{NL}(\lambda_0)$& $75.00$ $W^{-1}.km^{-1}$ & $64.82$ $W^{-1}.km^{-1}$ & $74.38$ $W^{-1}.km^{-1}$ \\
$A_{eff}$ & $2.70$ $\mu m^2$ & $3.77$ $\mu m^2$ & $3.53$ $\mu m^2$ \\
\bottomrule
\end{tabular}
\end{table}

Table~\ref{table:parameter_values} details the dispersion coefficients \(\beta_2\), \(\beta_3\), and \(\beta_4\), along with nonlinear parameters and mode area at the respective ZDWs for \(d_c = 2.4\,\mu\text{m}\) and \(1.4\,\mu\text{m}\). As \(d_c\) decreases, \(|\beta_2|\) increases markedly, indicating stronger anomalous dispersion. The effective mode area shrinks from \(2.75\,\mu\text{m}^2\) to \(1.6\,\mu\text{m}^2\), and \(\gamma\) rises from \(72\) to \(124\,\text{W}^{-1}\text{km}^{-1}\). These trends highlight the amplified nonlinear effects achievable with tighter modal confinement.

\begin{table}[htbp!]
\centering
\caption{$\beta_n$, $A_{eff}$, $\gamma$ and $L$ Values of Structures with Varying $d_c$ at $\lambda_0$}
\label{table:parameter_values}
\begin{tabular}{lcc}
\toprule
\textbf{Optical Parameters} & \textbf{$d_c=2.4$ $\mu m$} & \textbf{ $d_c=1.4$ $\mu m$} \\
\midrule
$\beta_2$ [ps\textsuperscript{2}/km] & $-15.07$ & $-46.2$ \\
$\beta_3$ [ps\textsuperscript{3}/km] & $0.0437$ & $0.0311$ \\
$\beta_4$ [ps\textsuperscript{4}/km] & $-0.0954$ & $-0.00665$ \\
$A_{\text{eff}}$ [$\mu$m\textsuperscript{2}] & $2.75$ & $1.6$ \\
$\gamma$ [W\textsuperscript{-1} km\textsuperscript{-1}] & $72$ & $124$ \\
\bottomrule
\end{tabular}
\end{table}

To assess contamination sensitivity, we modeled IPA infiltration by adjusting the refractive index of the air holes. As shown in Fig.~\ref{fig:disp_hexipa} and Fig.~\ref{fig:disp_circipa}, IPA significantly alters the dispersion profile and shifts \(\lambda_0\). These shifts must be accounted for in sensor design, as IPA-induced refractive index changes can degrade or enhance optical responses depending on the target operating wavelength.

\begin{figure}[htbp]
\centering
\includegraphics[width=\textwidth]{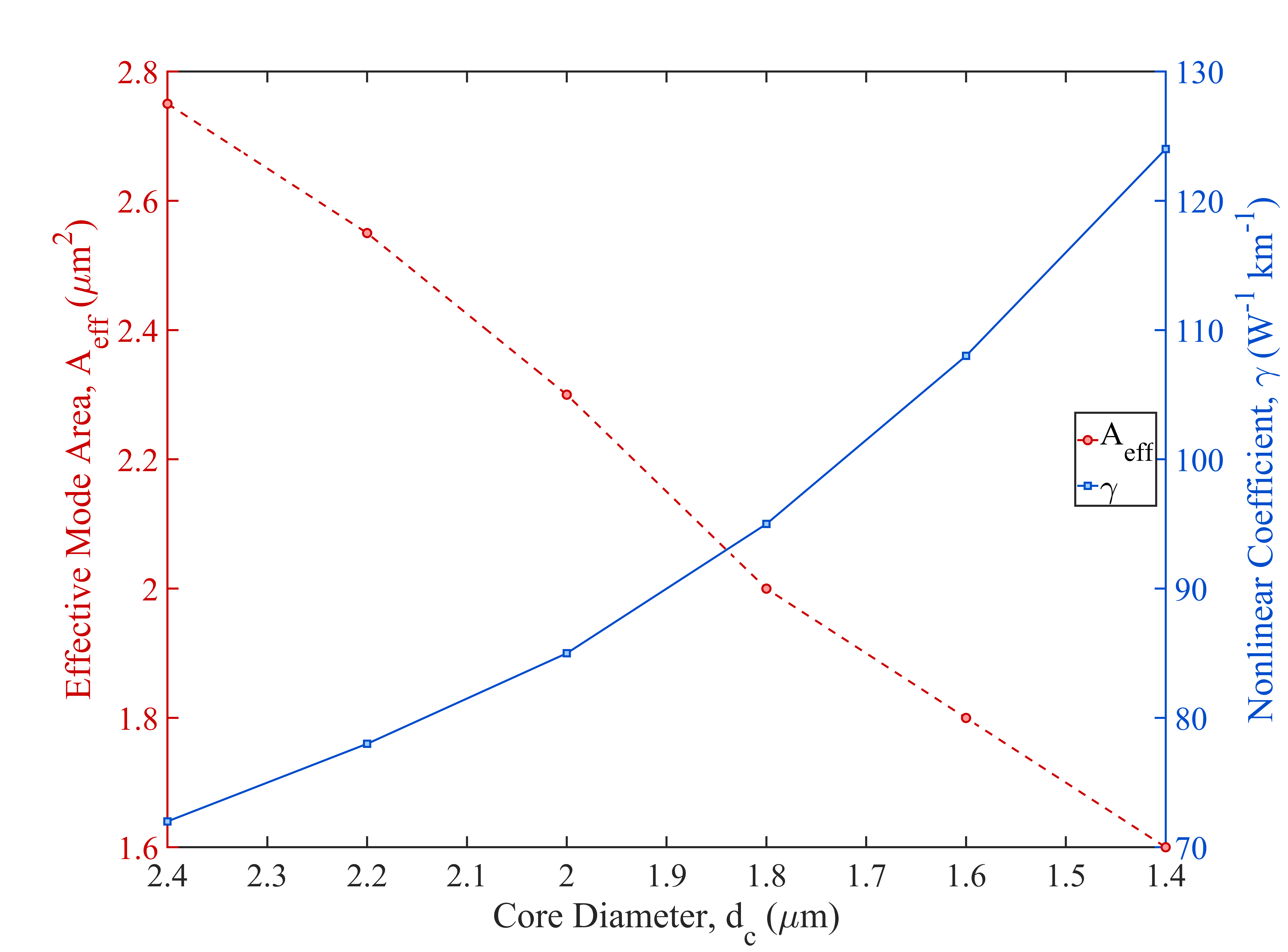}
\caption{Variation of effective mode area $A_{\text{eff}}$ (left axis, red dashed line with circles) and nonlinear coefficient $\gamma$ (right axis, blue solid line with squares) as a function of core diameter $d_c$. The results show that reducing $d_c$ decreases $A_{\text{eff}}$ and correspondingly enhances $\gamma$, consistent with the inverse proportionality $\gamma \propto 1/A_{\text{eff}}$.}
\label{fig:Aeff_gamma}
\end{figure}

\begin{figure}[htbp]
\centering
\includegraphics[width=\textwidth]{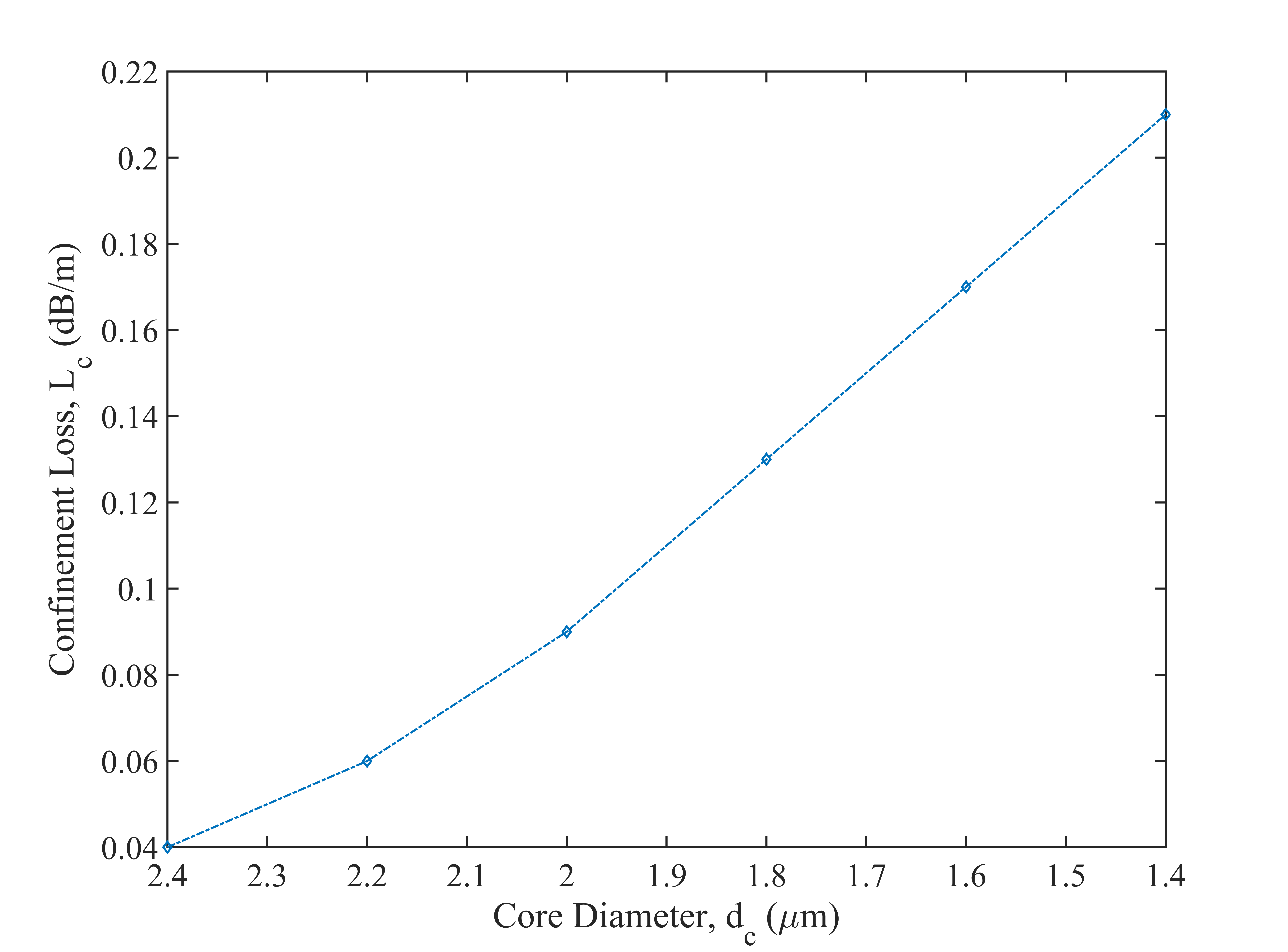}
\caption{Confinement loss $L_c$ as a function of core diameter $d_c$. Smaller core diameters result in significantly higher leakage loss, demonstrating the trade-off between tighter confinement (enhanced $\gamma$) and increased attenuation.}
\label{fig:conf_loss}
\end{figure}

In line with the properties highlighted in the abstract, we extended our graphical analysis to include the effective mode area ($A_{\text{eff}}$), nonlinear coefficient ($\gamma$), and confinement loss ($L_c$). Fig.~\ref{fig:Aeff_gamma} illustrates the strong inverse relationship between $A_{\text{eff}}$ and $\gamma$, where decreasing $d_c$ enhances nonlinear performance by reducing modal area. However, this improvement comes at the expense of higher confinement loss, as depicted in Fig.~\ref{fig:conf_loss}, which shows an exponential rise in $L_c$ with decreasing $d_c$. These results directly address the reviewer’s concern and confirm the interplay of geometry, nonlinearity, and attenuation. The combined trends emphasize the critical balance required in PCF design: optimizing $d_c$ for nonlinear efficiency while maintaining acceptable loss for practical applications such as supercontinuum generation and sensing.

Overall, our analysis confirms that the optical properties of PCFs are highly sensitive to geometric modifications. Core diameter \(d_c\) governs dispersion, attenuation, confinement loss, group index, and nonlinearity in a tightly coupled manner. Validation against experimental data confirms the accuracy of our simulations. The modeling of IPA infiltration demonstrates the importance of environmental robustness in real-world PCF deployment. These insights are critical for engineering PCFs that meet the stringent demands of contemporary photonic applications in nonlinear optics, telecommunication, and sensing.

\FloatBarrier
\section{Discussion}

The simulation results presented in this study reveal a complex interplay between structural geometry, environmental refractive index, and optical performance in photonic crystal fibers (PCFs). Several critical trends and trade-offs emerge that inform the optimization of PCF architectures for applications in nonlinear photonics and optical sensing.

A key observation is that reducing the core diameter \(d_c\) produces a substantial blue-shift in the zero-dispersion wavelength (ZDW). This shift results from enhanced waveguide dispersion, which becomes increasingly dominant over material dispersion as \(d_c\) is reduced into the sub-micron regime. Our simulations show that \(\lambda_0\) shifts from \(791\,\text{nm}\) to \(646\,\text{nm}\) as \(d_c\) decreases from \(2.4\,\mu\text{m}\) to \(1.4\,\mu\text{m}\). However, this spectral tunability is accompanied by increases in both attenuation \(\alpha\) and confinement loss \(L_c\), as indicated in Table~\ref{table:attenuation} and Fig.~\ref{fig: PCFlcall}. The trade-off between dispersion engineering and modal confinement must therefore be carefully managed.

The nonlinear coefficient \(\gamma\) exhibits a strong inverse dependence on the effective mode area \(A_{\text{eff}}\). As \(d_c\) decreases, \(A_{\text{eff}}\) contracts due to tighter field confinement, which in turn enhances the nonlinear coefficient significantly—from \(72\,\text{W}^{-1}\text{km}^{-1}\) to \(124\,\text{W}^{-1}\text{km}^{-1}\). This increase in \(\gamma\) is beneficial for high-efficiency supercontinuum generation, soliton dynamics, and other nonlinear effects. However, the accompanying increase in loss must be accounted for, particularly when designing for long-distance or high-power transmission regimes.

The role of geometric topology is also non-negligible. Both circular and hexagonal air-hole configurations exhibit qualitatively similar behavior in terms of dispersion and nonlinearity, but hexagonal lattices offer superior performance in several respects. For the same pitch, hexagonal claddings yield better confinement, lower \(A_{\text{eff}}\), and slightly higher \(\gamma\). This improvement is attributed to their denser air-hole packing and higher structural symmetry, which enhances the index contrast and modal confinement without increasing fabrication complexity.

Environmental robustness is another dimension explored in this study through the simulation of isopropyl alcohol (IPA) infiltration. IPA raises the cladding index from approximately \(1.00\) (air) to \(1.377\), which reduces the index contrast with the silica core. This leads to a red-shift in the ZDW by approximately \(20\)--\(35\,\text{nm}\) and increases leakage losses, particularly near the ZDW region. These results underscore the necessity of mitigating environmental contamination in field-deployed PCFs or, alternatively, using this sensitivity for refractive index-based sensing. In practical terms, protective coatings or hermetic sealing may be essential for applications requiring long-term stability.

Validation against experimental results—specifically the Thorlabs NL-2.3-790-02 fiber—demonstrates the accuracy and reliability of our computational models. The observed deviations in \(\lambda_0\), \(\gamma\), and \(A_{\text{eff}}\) are all under \(4\%\), which confirms the predictive capability of our full-vector FDTD and PWE methods when appropriately meshed and calibrated. This level of agreement strengthens confidence in the simulation framework and its utility for predictive fiber design. 

Clarify the fabrication of solid-core PCFs with hexagonal cladding is relatively straightforward using the stack-and-draw method. However, reducing the core diameter to sub-micron values introduces increased sensitivity to fabrication tolerances and imperfections.

Taken together, these results advocate for a modular, tunable design framework where \(d_c\), pitch \(\Lambda\), and cladding index are treated as primary parameters to engineer fiber response. This framework facilitates the development of application-specific PCFs, such as telecom-band supercontinuum sources, compact nonlinear modules, and contamination-aware refractive index sensors. The ability to simulate and optimize these structures with high fidelity provides a robust pathway toward the next generation of high-performance photonic crystal fibers.

The proposed PCF design is particularly suited for supercontinuum generation (SCG). Its tunable zero-dispersion wavelength (646--791~nm) aligns with common femtosecond pump sources such as Ti:Sapphire (~800~nm) and Yb-doped fiber lasers (~1060~nm), while the high nonlinear coefficient (\(\gamma\) up to \(124 \,\text{W}^{-1}\text{km}^{-1}\)) enables efficient spectral broadening at low pulse energies. Compared to existing PCFs (e.g., Thorlabs NL series), our design offers the combined advantages of higher nonlinearity and greater robustness against degradation from contaminants such as isopropyl alcohol (IPA), improving both efficiency and practical applicability.

The systematic evaluation and optimization of PCF structures, supported by both \cite{thorlabs:pcf} and theoretical analyses, enable the development of highly efficient and adaptable photonic devices. The ability to manipulate the optical properties of PCFs through structural modifications and material choices is pivotal for advancing the field of photonics and achieving breakthroughs in various technological applications.

In the context of dispersion engineering in our study by adjusting the core diameter $(d_c)$, we can precisely tune the ZDW, as shown in Fig. \ref{fig:disp}. This capability is essential for applications including pulse compression, supercontinuum generation, and dispersion compensation in optical communication systems. Also, the linear impact of decreasing $(d_c)$ on the ZDW provides a straightforward approach to achieving the desired dispersion characteristics.

\begin{figure}[htbp]
\centering
\includegraphics[width=\textwidth]{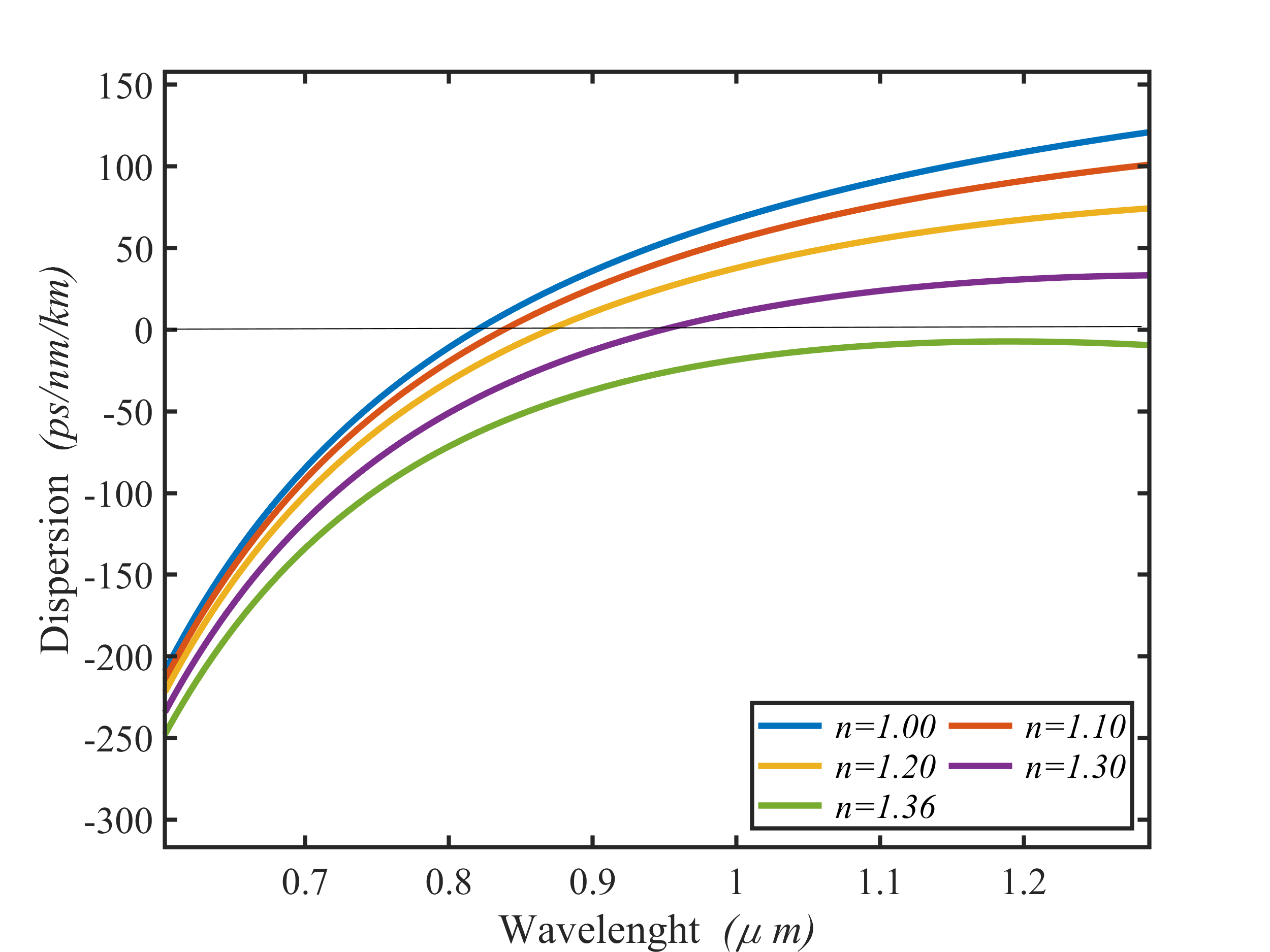}
\caption{Dispersion behavior and ZDW for circular rods with different IPA fillings of PCF-1.}
\label{fig:disp_hexipa}
\end{figure}

\begin{figure}[htbp]
\centering
\includegraphics[width=\textwidth]{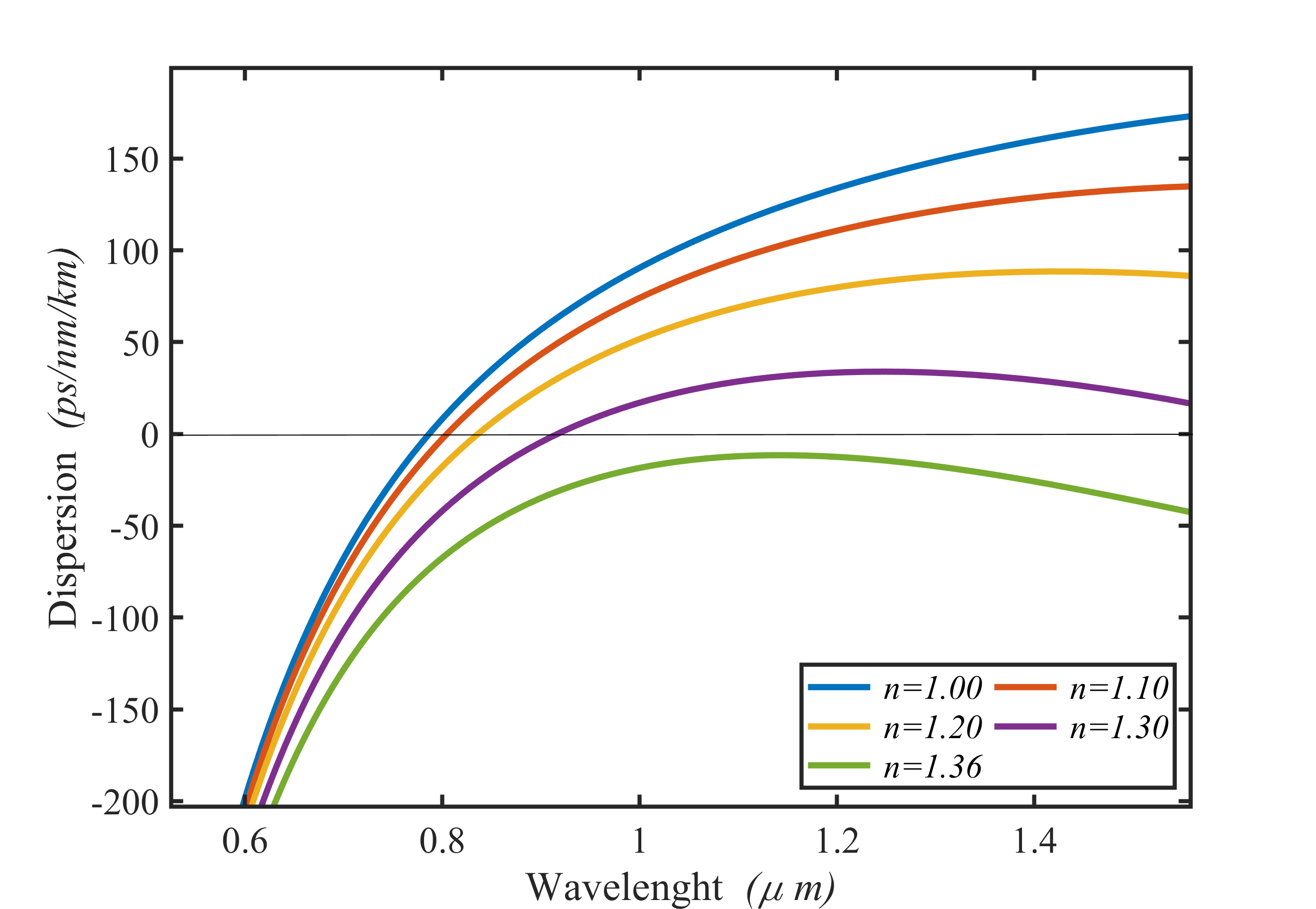}
\caption{Dispersion behavior and ZDW for circular rods with different IPA fillings of PCF-2.}
\label{fig:disp_circipa}
\end{figure}

 Our analysis, depicted in Fig. \ref{fig: PCFlcall}, indicates that attenuation increases with a decrease in $(d_c)$. This balance between dispersion management and confinement efficiency highlights the critical need for optimization in PCF design. Balancing low dispersion with minimal attenuation is essential for maximizing the performance of PCFs in various applications. The variations in group index \(n_g\) with different ($d_c$)'s, shown in Fig. \ref{fig: PCFngall}, are crucial for controlling pulse propagation and group velocity dispersion. This is particularly relevant for high-speed optical communication and precision sensing technologies. Understanding and controlling the group index behavior enables the design of PCFs with tailored propagation characteristics suited to specific applications. The comprehensive dataset provided in Tables \ref{table:parameter_values} and \ref{table:dispersion_values} offers valuable insights into the relationships between structural parameters and optical properties. This data is essential for the further refinement and optimization of PCF designs.  We can achieve a wide range of optical properties tailored to specific application requirements by systematically varying parameters such as ($d_c$) and lattice pitch. Filling the PCF structures with isopropyl alcohol (IPA) significantly modified the dispersion characteristics, as illustrated in Figs. \ref{fig:disp_hexipa} and \ref{fig:disp_circipa}. This demonstrates the impact of IPA contamination on the performance of PCFs. IPA modifies the refractive index contrast within the fiber, influencing the ZDW and the overall dispersion profile. As a result, this unwanted condition adversely affects the quality of supercontinuum generation. Understanding these effects is crucial for maintaining optimal performance in sensor applications where such contaminants may be present.

\FloatBarrier
\section{Conclusion}

The findings of this study highlight the versatility and tunability of PCFs, paving the way for innovative advancements in photonic technologies. Future research will focus on several key areas, such as investigating new materials with unique refractive indices and nonlinear properties to further enhance the performance and functionality of PCFs. Additionally, exploring more complex PCF geometries, such as multi-core and hybrid structures, will aim to achieve even greater control over optical properties.

This study contributes novelty and originality to the literature by systematically analyzing the impact of core diameter and material modifications on PCF performance. The comprehensive evaluation of dispersion, attenuation, and group index across different structural configurations provides a valuable reference for future research. Moreover, the investigation into the effects of IPA contamination on PCF properties adds a practical dimension to the study, highlighting real-world considerations for sensor applications. The insights gained from this research will inform the design and optimization of next-generation photonic devices, ensuring their robustness and efficiency in diverse technological applications.

In conclusion, tuning the core diameter allows precise control of ZDW 
\((646\text{--}791~\text{nm})\), enabling applications such as supercontinuum generation. 
The enhanced nonlinear coefficient \(\gamma\) (up to \(124~\text{W}^{-1}\text{km}^{-1}\)) 
supports efficient broadband continuum sources. Compared to existing PCFs, the proposed design 
offers a better balance between dispersion and nonlinearity. Fabrication feasibility is high, 
though sub-micron cores increase complexity.

\section{Acknowledgement}

This study was supported by the Scientific and Technological Research Council of Turkey (TUBITAK) under Project No. 118E954. 

\bibliographystyle{unsrtnat}
\bibliography{bibi.bib}

\end{document}